# Anomalously Strong Size Effect on Thermal Conductivity of Diamond Microparticles

Yufeng Wang[1] and Bo Sun[1]

*Tsinghua Shenzhen International Graduate School, Tsinghua University, Shenzhen 518055, China.*

(*Electronic mail: sun.bo@sz.tsinghua.edu.cn)

(Dated: 2 April 2024)

Diamond has the known highest thermal conductivity of around 2000 Wm$^{-1}$ K$^{-1}$ and is therefore widely used for heat dissipation. In practical applications, synthetic diamond microparticles are usually assumed to have similar thermal conductivity to that of bulk diamond because the particle size is larger than theoretical phonon mean free path so that boundary scattering of heat-carrying phonons is absent. In this report, we find the thermal conductivity of diamond microparticles anomalously depends on their sizes. Thermal conductivity of diamond microparticles increases from 400 Wm$^{-1}$ K$^{-1}$ to 2000 Wm$^{-1}$ K$^{-1}$ with the size growing from 20 µm to 300 µm. We attribute the abnormally strong size effect to the long-range defects during the growth process based on analysis of point defects, dislocations, and thermal penetration depth dependence of thermal conductivity. Our results play a vital role in the design of diamond composites and in the improvement of thermal conductivity of synthetic diamonds.

## I. INTRODUCTION

With the decreasing size of modern semiconductor devices and the increasing demand for high-power electronics, heat dissipation has become a major challenge limiting the performance of electronic devices. An effective thermal management system usually needs a material with high thermal conductivity that can efficiently transfer heat out of hot spots. Among all materials, diamond has the highest thermal conductivity, which is 2400 Wm$^{-1}$ K$^{-1}$ reported in high-purity single crystalline diamond[1,2], making it the best thermal conductor in thermal management applications.

In industrial applications, diamond-reinforced composites are expected to be excellent thermal management materials and have attracted considerable attention[3], as diamond particles have high thermal conductivity and metal or polymer matrix has required mechanical properties. However, thermal conductivity of the reported composites is much lower than expected[4–7]. Past work mainly focused on preparation techniques and interface thermal conductance between diamond particles and the matrix to improve composites' thermal conductivity, where thermal conductivity of diamond microparticles with different sizes was assumed to be the same as that of bulk diamond in the design of the composites[8–10]. This assumption is supported by first-principles calculations that phonon mean free paths (MFPs) below 4 µm account for 90% of the thermal conductivity in diamonds[11–13]. In other words, the thermal conductivity of synthetic single-crystal diamond microparticles hardly changes with size and should be similar to that of bulk diamond.

Here, we used time-domain thermoreflectance (TDTR)[14] to measure diamond microparticles with sizes ranging from 20 µm to 300 µm and found a strong size effect on thermal conductivity of diamond microparticles. It is surprising that the thermal conductivity of 20 µm diamond particles is only around 400 Wm$^{-1}$ K$^{-1}$, which is much lower than the thermal conductivity of bulk diamond while their size is much larger than the first-principles calculated MFPs. The thermal conductivity of diamond microparticles increases with their sizes. When the size reaches 300 µm, the thermal conductivity of the diamond is 2000 Wm$^{-1}$ K$^{-1}$ which is a widely accepted value for high-quality diamond.

## II. EXPERIMENTAL METHODS

Four groups of diamond particles from 20 µm to 300 µm were synthesized using the typical high pressure and high temperature (HPHT) process[15,16]. The 2-5 µm seed diamonds and high-purity graphite powders as carbon sources were used to grow diamond particles. After mixing the raw materials, they were pressed into a disk to fit the shape of the HPHT apparatus. The size of diamond microparticles was solely controlled by growth time with other preparation conditions kept the same.

Thermal conductivity of diamond microparticles was measured by TDTR. The diamond particles were put onto carbon tape and then coated with 80 nm aluminum as TDTR transducer using an electron beam evaporation system. A 785 nm pump laser modulated by an electro-optic modulator with adjustable modulation frequency was absorbed by aluminum film and converted to heat. The time-delayed probe laser was focused on the same spot and partially reflected by aluminum film. The reflected probe laser containing a signal of temperature oscillation via the change of optical reflectance was detected by a photodetector and lock-in amplifier. The 6 µm, 12 µm, and 24 µm $1/e^2$ diameter of the Gaussian laser beams and the modulation frequencies of 10.1 MHz, 4.6 MHz, 1.01 MHz, and 0.47 MHz were used in the measurement. We determined thermal conductivity of diamond microparticles by fitting the time dependence of experimental results with a thermal diffusion model[14].

The optical microscope images were taken with Olympus SZX16. The scanning electron microscope (SEM) and electron backscatter diffraction (EBSD) were measured by Bruker QUANTAX CrystAlign 400i. The nitrogen content of diamond microparticles was characterized by Raman spec-troscopy using an 1800 gr mm$^{-1}$ grating and a 532 nm laser with a power of 5.13 mW.

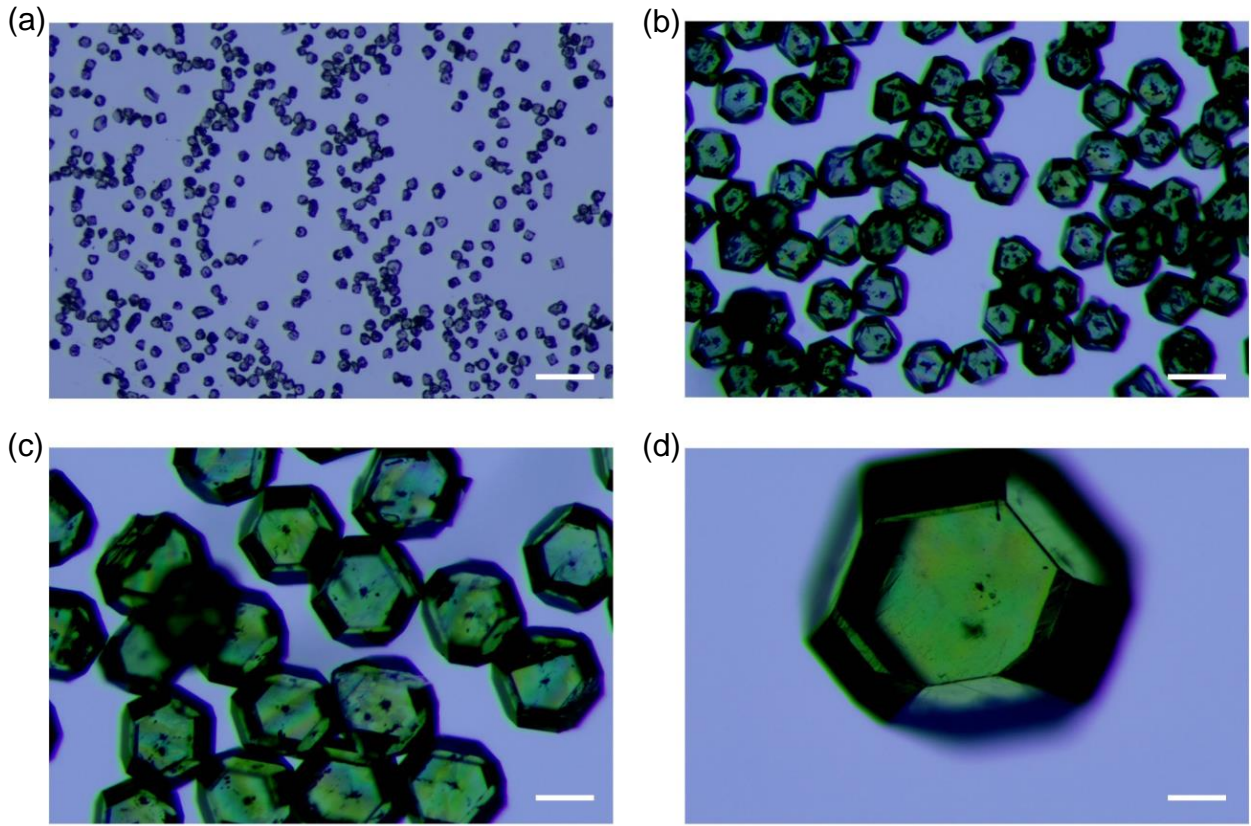

FIG. 1. Optical microscope images of the (a) 20 µm, (b) 60 µm, (c) 130 µm, (d) 300 µm diamond particles. The scale bar is 100 µm.

### III. RESULTS AND DISCUSSION

The optical microscope images (Fig. 1) show that the diamond microparticles are uniform in size and have smooth surfaces. We determined the size of diamond microparticles by the statistics of edge length on its surface and the sizes of four groups of samples are 20 µm, 60 µm, 130 µm, and 300 µm. We can see that there are opaque areas of seed diamonds in the center of the diamond microparticles by adjusting the transmitted and incident light in the optical microscope. In order to ensure that the diamond microparticles are single crystals, the EBSD experiment was performed. The SEM image and EBSD maps of a 20 µm diamond particle are shown in Fig. 2. We found that the surface of the smallest diamond microparticles already has a uniform crystal orientation. Thus, we can infer that all synthesized diamond microparticles in this study are single crystals and the opaque areas shown in the optical microscope do not affect the crystal structure of synthetic diamonds.

Thermal conductivity of diamond microparticles was measured by TDTR, which is a pump-probe optical technique with a spatial resolution on the order of 10 µm and adjustable heat penetration depth. Therefore, we can measure thermal conductivity of an individual diamond particle at different depths. For 20 µm diamonds, we only used 6 µm $1/e^2$ diameter of the Gaussian beam and the modulation frequency of 10.1 MHz, as limited by their size and heat diffusion length. For diamond microparticles with other sizes, we used different diameters of the Gaussian beam and modulation frequencies in TDTR measurements. The thermal conductivity results are the same when the diameter of the Gaussian beam is varied, but they are different when the modulation frequency changes, which will be discussed later. The measured maximum thermal conductivity and *in situ* micrograph of diamond particles with various sizes is shown in Fig. 3. Thermal conductivity of diamond microparticles increases with the size of the diamond, from 400 Wm$^{-1}$ K$^{-1}$ to 2000 Wm$^{-1}$ K$^{-1}$, which is totally different from conventional wisdom that thermal conductivity of diamond microparticles should be close to 2000 Wm$^{-1}$ K$^{-1}$, the value of bulk diamond.

One possibility for this anomalous size-dependent thermal conductivity is that point defect concentrations vary in different sized samples. Since nitrogen is the most common impurity in HPHT diamonds[17], We characterized the nitrogen content in our diamond microparticles using Raman spectroscopy (Fig. 4a). The peak of 1332.5 cm$^{-1}$ is the first-order diamond

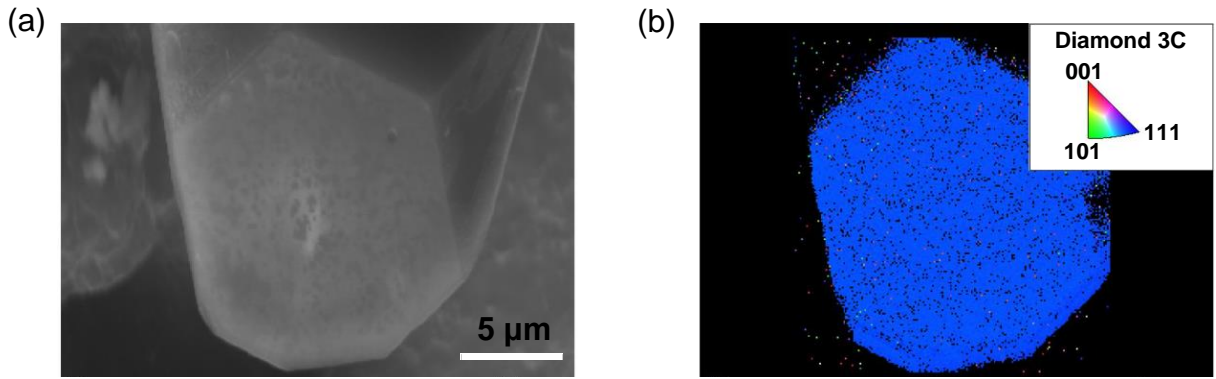

FIG. 2. (a) SEM image of a 20 µm diamond particle. (b) EBSD maps of the diamond particle. The inset is crystal orientation of diamond 3C.

Raman line and the width of the Raman line is proportional to the nitrogen content in diamond[18]. We found that the nitrogen content decreased slightly from 365 ppm to 222 ppm as the diamond microparticle size increased. However, previous first-principles calculation results[19] on thermal conductivity of diamond affected by nitrogen impurities showed that the difference in nitrogen content between 365 ppm and 222 ppm has a limited effect on thermal conductivity (Fig. 4b). In addition, past experiments about millimeter-scale diamonds reported a slight decrease in thermal conductivity with increasing nitrogen contents[20,21], suggesting that nitrogen content at this level is not a key factor causing the size effect. Other point defects, such as vacancies, isotope, and boron impurity, should have similar concentrations in different diamond microparticles because of the same raw material and growth conditions in the HPHT process. Hence, the size effect cannot be fully explained by point defects in diamond microparticles.

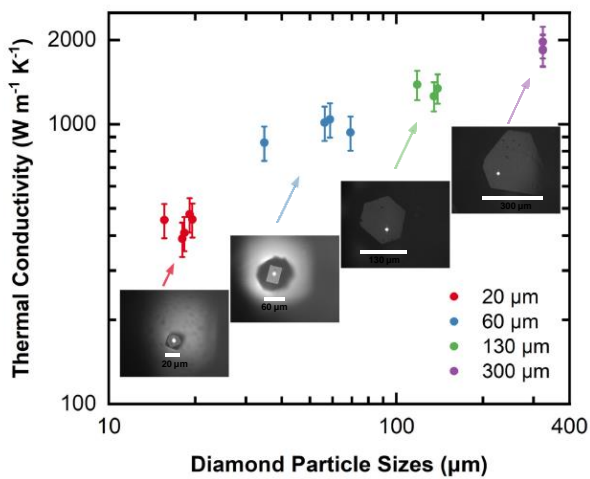

FIG. 3. Thermal conductivity and micrograph of diamond microparticles. The bright laser beam on the surface of sample is the measurement position of TDTR.

Dislocations may be one of the possibilities because the dislocation density in diamonds increases with the decrease of diamond crystal size[22]. However, homoepitaxial HPHT single-crystal diamond has typically low dislocation density (between $10^4$ cm$^{-2}$ and $10^6$ cm$^{-2}$)[23]. Past experiment has shown that dislocation density with $10^{10}$ cm$^{-2}$ has a limited effect on thermal conductivity in indium nitride at room temperature[24]. For the size effect of thermal conductivity, therefore, dislocation density is not an important parameter.

Another possibility for this anomalous size effect is that the first-principles calculated phonon MFPs is underestimated in diamond. While first-principles calculation has been developed to be an accurate method to obtain phonon MFP, experiment results are always required to determine phonon MFP. TDTR is a powerful technique to accurately determine phonon MFP because modulation frequency ($f$) can adjust the thermal penetration depth, $d = \sqrt{\frac{\Lambda}{\pi C f}}$, where $\Lambda$ is the thermal conductivity and $C$ is the volumetric heat capacity[25]. As $f$ increases, the length scales of the temperature profile will be shorter than the long MFP of diamond phonons, which will lead to partial phonon non-equilibrium transport in the measurement of thermal conductivity, resulting in reduced thermal conductivity[25,26]. The previous calculations showed the apparent thermal conductivity could be approximated by assuming an additional boundary scattering at a characteristic length $2d$ when the distribution of phonon MFPs is sufficiently wide[27,28]. The phonon MFPs of diamond span at least two orders of magnitude[11], so the characteristic length $2d$ can be used to approximately estimate the distribution of phonon MFPs in diamond. Fig. 5a shows thermal conductivity at different characteristic lengths $2d$ and cumulative thermal conductivity of bulk diamond as a function of the phonon MFPs[11]. We assume that the cumulative maximum thermal conductivity is 2000 Wm$^{-1}$ K$^{-1}$. For one 300 µm diamond particle, the thermal conductivity we measured is 1562 Wm$^{-1}$ K$^{-1}$ when $f$ of 10.1 MHz is used which means $2d$ is only around 10.6 µm shorter than long MFP of diamond phonons. As $2d$ increases to 36.2 µm, the measured thermal conductivity of the diamond gradually increases to 1841 Wm$^{-1}$ K$^{-1}$, which means that around 15% of the ther-mal conductivity is contributed by phonon MFPs larger than 10.6 μm. However, upon consideration of experimental uncertainties, our findings exhibit a small deviation from previous past first-principles calculations where phonon MFPs below 4 μm account for 90% of the thermal conductivity in diamonds[11–13].

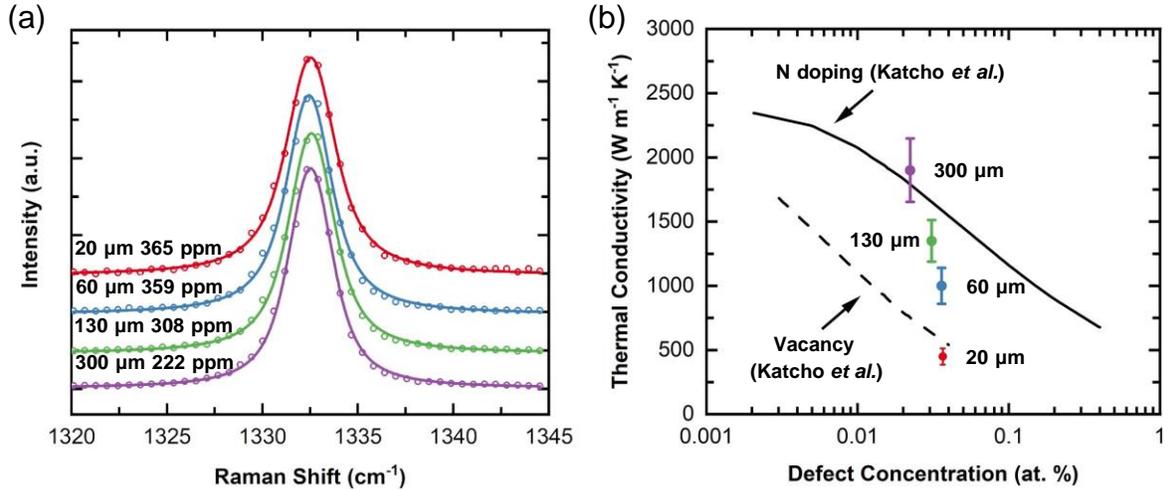

FIG. 4. (a) Raman spectroscopy of diamond microparticles. (b) Thermal conductivity affected by point defects in our experiment and first-principles calculation results[19].

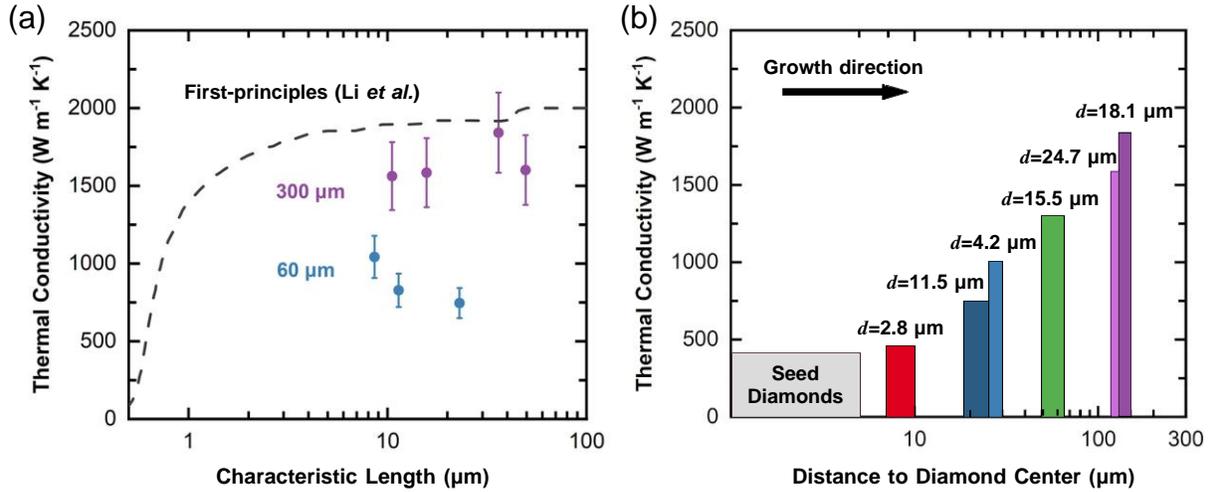

FIG. 5. (a) Thermal conductivity of 60 μm and 300 μm diamond particles as a function of $2d$. The dashed line is the phonon MFPs of diamond calculated from first-principles[11]. (b) Thermal conductivity as a function of distance to diamond center. The width of each bar represents the heat penetration depth.

When we tested the 300 μm diamond particle, we observed the thermal conductivity reduced to 1601 Wm$^{-1}$ K$^{-1}$ with a $f$ of 0.47 MHz ($2d$ of 49.4 μm) suggesting that a hidden defect exists deep within the diamond. Furthermore, in a 60 μm diamond particle, thermal conductivity even monotonically decreases with increasing thermal penetration depth (Fig. 5a). We infer that the hidden defect has a more significant impact on thermal conductivity as it gets closer to the growth core. We show the relationship between thermal conductivity and distance to the diamond center shown in Fig. 5b. This size effect suggests that there may be long-range defects spanning as long as 150 μm during the diamond growth process, in which the structure and properties of the diamond are improved with size increasing. The long-range defects probably come from inhomogeneous strain, whose presence has been observed by experimental X-ray rocking curves in previous reports[29,30]. The strain can cause unprecedented structural detail and complexity within diamonds, such as stacking-disordered structure, nanotwins, and nanostructures, which were observed in diamonds by aberration-corrected high-resolution transmis-sion electron microscopy (TEM)[31]. The phonons will scatter at the boundary of nanostructures and the phonon MFPs will be limited to the size of the complex structure. Therefore, we infer that the anomalously strong size effect on thermal conductivity is affected by the long-range defects.

## IV. CONCLUSION

We found the thermal conductivity of synthetic diamond microparticles, which is always assumed around 2000 Wm$^{-1}$ K$^{-1}$ in past applications, changes with growth size. In the smallest 20 μm particles, the thermal conductivity is 400 Wm$^{-1}$ K$^{-1}$, only one-fifth of what is predicted from first-principles calculations. We infer the size effect arises from the long-range defects during growth progress based on the anomalous dependence of $d$ in the measurement of thermal conductivity. The size effect will play an important role in the design of high thermal conductivity composites and in thermal management systems using synthetic diamond. The thermal conductivity is sensitive to diamond quality and therefore our results will provide guidance for improvement of diamond synthetic processes.

## ACKNOWLEDGMENTS

We acknowledges support from NSFC-ISF Joint Scientific Research Program under Grant No. 52161145502, National Science Foundation of China under Grant No. 12004211, Shenzhen Science and Technology Program grants RCYX20200714114643187 and WDZC20200821100123001, Tsinghua Shenzhen International Graduate School grants QD2021008N and JC2021008.

## DATA AVAILABILITY STATEMENT

The data that support the findings of this study are available from the corresponding author upon reasonable request.